\documentclass[aps,prx,amsmath,superscriptaddress,eqsecnum,twocolumn,nofootinbib,longbibliography]{revtex4-2}
\pdfoutput=1
\usepackage[utf8]{inputenc}

\usepackage{times}
\usepackage{microtype}

\usepackage{graphicx}
\usepackage[dvipsnames]{xcolor}
\usepackage{amssymb,amsmath,mathtools}
\usepackage{tensor}
\usepackage{adjustbox}
\usepackage{booktabs}
\usepackage{tabularx}
\usepackage{multirow,makecell}
\usepackage{enumitem}
\usepackage{etoolbox}
\usepackage{mathtools}
\usepackage{titlesec}
\usepackage{euscript} 
\usepackage{fontawesome}
\usepackage{pifont}
\newcommand{\cmark}{\text{\ding{51}}}
\newcommand{\xmark}{\text{\ding{55}}}

\makeatletter
\newcommand*{\defeq}{\mathrel{\rlap{%
			\raisebox{0.3ex}{$\m@th\cdot$}}%
		\raisebox{-0.3ex}{$\m@th\cdot$}}%
	=}
\newcommand*{\eqdef}{=\mathrel{\rlap{%
			\raisebox{0.3ex}{$\m@th\cdot$}}%
		\raisebox{-0.3ex}{$\m@th\cdot$}}%
}
\makeatother

\makeatletter
\g@addto@macro\bfseries{\boldmath}
\makeatother

\usepackage{etoolbox}
\usepackage{caption}

\makeatletter
\def\thesubsection{\thesection.\arabic{subsection}}
\def\p@subsection{}
\makeatother

\titleformat{\section}[hang]
{\normalfont\fontsize{10.5pt}{12pt}\selectfont\bfseries}
{\thesection.}{0.75em}
{\raggedright}

\titleformat{\subsection}[hang]
{\normalfont\normalsize\bfseries}
{\hspace{1em}\thesubsection.}{0.5em}
{\hspace{0pt}\raggedright}

\titleformat{\subsubsection}[hang]
{\normalfont\normalsize\bfseries}
{\thesubsubsection.}{1em}{\raggedright}

\titlespacing*{\section}{0pt}{4mm}{1mm}
\titlespacing*{\subsection}{0pt}{3mm}{0.5mm}
\titlespacing*{\subsubsection}{0pt}{2.25mm}{0.25mm}

\apptocmd{\appendix}{%
	\makeatletter
	\renewcommand{\thesubsection}{\thesection.\arabic{subsection}}%
	\makeatother
	\titleformat{\section}[display]%
	{\normalfont\normalsize\bfseries}%
	{Appendix \thesection}%
	{0.3em}%
	{\raggedright}%
	\titlespacing*{\section}{0pt}{3mm}{2.2mm}%
}{}{}

\makeatletter
\def\l@f@section{%
	\addpenalty{\@secpenalty}%
	\addvspace{0.60em plus 0.10em}%
}
\def\l@@sections#1#2#3#4{%
	\begingroup
	\everypar{}%
	\set@tocdim@pagenum\@tempboxa{#4}%
	\global\@tempdima\csname tocdim@#2\endcsname
	\@nameuse{l@f@#2}%
	\setbox\z@\vbox{%
		\hsize\linewidth
		\leftskip\csname tocleft@#2\endcsname\relax
		\dimen@\csname tocleft@#1\endcsname\relax
		\parindent-\leftskip\advance\parindent\dimen@
		\rightskip\tocleft@pagenum plus 1fil\relax
		\skip@\parfillskip\parfillskip\z@
		\let\numberline\numberline@@sections
		\ignorespaces#3\unskip\par
	}%
	\setbox\tw@\vbox to \ht\z@{%
		\vfil
		\hb@xt@\tocleft@pagenum{\hfil\unhbox\@tempboxa}%
		\vfil
	}%
	\noindent
	\rlap{\hb@xt@\linewidth{\hfil\box\tw@}}%
	\box\z@
	\par
	\expandafter\aftergroup\csname tocdim@#2%
	\expandafter\endcsname
	\expandafter\endgroup
	\the\@tempdima\relax
}
\makeatother

\usepackage{scalerel}
\usepackage{tikz}
\usetikzlibrary{svg.path}
\definecolor{orcidlogocol}{HTML}{A6CE39}
\tikzset{
	orcidlogo/.pic={
		\fill[orcidlogocol] svg{M256,128c0,70.7-57.3,128-128,128C57.3,256,0,198.7,0,128C0,57.3,57.3,0,128,0C198.7,0,256,57.3,256,128z};
		\fill[white] svg{M86.3,186.2H70.9V79.1h15.4v48.4V186.2z}
		svg{M108.9,79.1h41.6c39.6,0,57,28.3,57,53.6c0,27.5-21.5,53.6-56.8,53.6h-41.8V79.1z M124.3,172.4h24.5c34.9,0,42.9-26.5,42.9-39.7c0-21.5-13.7-39.7-43.7-39.7h-23.7V172.4z}
		svg{M88.7,56.8c0,5.5-4.5,10.1-10.1,10.1c-5.6,0-10.1-4.6-10.1-10.1c0-5.6,4.5-10.1,10.1-10.1C84.2,46.7,88.7,51.3,88.7,56.8z};
	}
}
\newcommand\orcidlink[1]{\href{https://orcid.org/#1}{\mbox{\scalerel*{
				\begin{tikzpicture}[yscale=-1,transform shape]
					\pic{orcidlogo};
				\end{tikzpicture}
			}{X}}}}

\usepackage{fancyhdr}
\usepackage{soul}
\usepackage[normalem]{ulem}

\def\gP{\mathfrak{P}}

\def\eF{\EuScript{F}}
\def\eH{\EuScript{H}}
\def\eO{\EuScript{O}}
\def\cL{\mathcal{L}}
\DeclareMathOperator{\Tr}{Tr}

\definecolor{arxivlink}{RGB}{0,140,220}
\definecolor{doilink}{RGB}{10,30,160}

\usepackage{url,hyperref}
\hypersetup{colorlinks,linkcolor={blue!55!black},citecolor={red!50!black},urlcolor={doilink},breaklinks=true}

\begin{document}
	
	\title{Classical-quantum gravity as quantum gravity in disguise}
	
	\author{Masahiro Hotta}
	\email{hotta@tuhep.phys.tohoku.ac.jp}
	\affiliation{Department of Physics, Tohoku University, Sendai 980-8578, Japan}
	\affiliation{Leung Center for Cosmology and Particle Astrophysics, National Taiwan University, Taipei 10617, Taiwan (R.O.C.)}
	
	\author{Sebastian Murk\orcidlink{0000-0001-7296-0420}}
	\email{sebastian.murk@matfyz.cuni.cz}
	\affiliation{Faculty of Mathematics and Physics, Charles University, Ke Karlovu 3, 121 16 Praha 2, Czech Republic}
	\affiliation{Quantum Gravity Unit, Okinawa Institute of Science and Technology, 1919-1 Tancha, Onna-son, Okinawa 904-0495, Japan}
	
	\author{Daniel R.\ Terno\orcidlink{0000-0002-0779-0100}}
	\email{daniel.terno@mq.edu.au}
	\affiliation{School of Mathematical and Physical Sciences, Macquarie University, New South Wales 2109, Australia}
		
	\begin{abstract}
		Whether gravity must be quantized remains one of the biggest open problems in fundamental physics. Classical-quantum hybrid theories have recently attracted attention as a possible framework in which gravity is treated classically yet interacts consistently with quantum matter. Schemes based on completely positive dynamics satisfy most formal consistency requirements and enable a systematic treatment of quantum backreaction, but they also invite the question of whether the hybrid description is fundamental or instead an effective sector of a larger quantum theory. Here, we demonstrate that classical-quantum gravity based on completely positive dynamics admits an embedding into a fully quantum theory on an enlarged Hilbert space. As a complementary illustration, we consider a qubit interacting with a classical particle and demonstrate that the corresponding hybrid system violates angular momentum conservation despite rotational symmetry of the underlying equations of motion. This provides an explicit example of a fully closed, rotationally invariant classical-quantum system with completely positive dynamics that violates a conservation law.
	\end{abstract}
	
	\maketitle
	\thispagestyle{fancy}
	
	\section{Introduction}
	A consistent mathematical description of the mutual interaction between classical and quantum systems --- known as hybrid dynamics --- is thought of either as a computational tool, as part of a resolution of the quantum measurement problem, or as a fundamental theory if the gravitational field remains at least partially classical. Hybrid models must satisfy certain axioms, but their degree of compliance depends on the intended purpose of the model. In computational applications, a hybrid scheme only needs to remain valid for the duration of the investigated process, although full consistency is still desirable. In contrast, a hybrid model aiming to describe a fundamental theory of gravity must meet stricter requirements, while models developed for other purposes may not be applicable to processes that involve gravitation \cite{Boucher.Traschen:PhysRevD:1988,Terno:chapter:2025}.
	
	Approaches to classical-quantum (CQ) dynamics broadly fall into two categories: reversible (unitary) and irreversible (completely positive). The former aim to provide a mathematically consistent counterpart to both unitary quantum and classical Hamiltonian theories without introducing dissipation or diffusion terms. However, all known linear reversible schemes are generally inconsistent. Nonlinear evolution, such as that found in mean-field systems, also leads to inconsistencies \cite{Terno:chapter:2025}, although many of these have been resolved by recently developed Ehrenfest--Koopman methods within their domain of validity \cite{Martinez-Crespo.Tronci:EurPhysJPlus:2025}. The breakdown of hybrid dynamics happens when the true quantum Wigner dynamics of the subsystem that is approximated as classical develops interference patterns that extend over phase space \cite{Bauer.etal:MultiscaleModelSimul:2024}. Moreover, for the description to remain valid over a finite period of time, the classical system must possess some minimal uncertainty \cite{Barcelo.etal:PhysRevA:2012,Ahmadzadegan.Mann.Terno:PhysRevA:2016}.
	
	There is strong evidence that coupling a classical system to a generalized probability theory --- and not just quantum mechanics --- necessarily introduces irreversibility in the resulting linear theory \cite{Galley.Giacomini.Selby:Quantum:2023}. Recently, Oppenheim and collaborators proposed a completely positive (CP) hybrid scheme \cite{Oppenheim:PhysRevX:2023,Oppenheim.etal:NatCommun:2023} that satisfies most consistency axioms (summarized in Appendix~\ref{app:ss:consistency.requirements}) without breaking down at some finite time. This approach has been successfully applied to several scenarios requiring the backreaction of quantum fields on a classical background and has also provided potentially observable bounds on gravitationally induced decoherence. In this work, we further show that this CP hybrid framework admits a fully quantum realization on an enlarged Hilbert space. The relevant completion is not a canonical quantization of the classical phase-space variables; rather, the hybrid theory arises as a reduced/open description obtained from a unitary completion together with an operational restriction to a commutative sector.
	
	The CP evolution of the combined system ensures compliance with the technical axioms I--IV, VI, and VII (see Appendix~\ref{app:ss:consistency.requirements}), but it comes at a cost: it prevents full compliance with conservation laws, which are at the heart of axiom V. Indeed, it reduces the dimensionality of the manifold of steady states and can violate conservation laws. The mathematical formulation of CP dynamics in infinite-dimensional spaces is highly nontrivial, and incorporating Hamiltonian dynamics of the gravitational field adds further technical challenges. To isolate these conceptual issues, we focus on a toy model and leave the full coupled dynamics for future research. In particular, while open Lindbladian and collapse models are known to violate certain conservation laws \cite{Bassi.etal:RevModPhys:2013,Albert.Jiang:PhysRevA:2014}, we demonstrate this phenomenon in a closed CQ model with CP dynamics and relate it to Liouvillian asymptotics.
	
	The remainder of this article is structured as follows: In Section~\ref{sec:CQ.dynamics}, we briefly summarize the mathematical formalism used to describe the dynamical evolution of CQ hybrid systems and introduce the notation. In Section~\ref{sec:CP.fundamental.issues}, we discuss steady states and conservation laws in the framework of CP dynamics, along with related conceptual issues. In Section~\ref{sec:embedding.CQgravity.into.Qtheory}, we show that completely positive CQ dynamics can be embedded into a fully quantum theory on an enlarged Hilbert space. In Section~\ref{sec:toy.model}, we provide a toy model that violates conservation of angular momentum despite rotational invariance of the underlying equations of motion. Finally, in Section~\ref{sec:discussion.conclusions}, we summarize the physical implications of these results and outline possible avenues for future work.

	\section{Classical-quantum dynamics} \label{sec:CQ.dynamics}
	Before we consider the dynamics of a combined CQ system, we set the notation and detail the description of its two subsystems.
	
	The classical system on its own is described by a phase space $\mathfrak{P}$, which is a $2n$-dimensional Poisson manifold with local coordinates given by positions and momenta, $z=(q,p)$. We use the measure $dz$ without specifying its explicit coordinate form. The states of the classical system are given by the Liouville probability density $\varrho(z)$. For a closed system, the evolution is generated by some Hamiltonian $H_\text{cl}$. A precisely known classical trajectory is represented by a distribution $\varrho(z,t)=\delta\big(z-z_0(t)\big)$.
	
	A quantum system is described on a separable Hilbert space $\eH$. Information about its state is contained in a positive semidefinite, trace-class, trace-one density operator $\rho$.
	
	Following Refs.~\cite{Oppenheim:PhysRevX:2023,Oppenheim.etal:NatCommun:2023}, we describe the state of the CQ system using a density operator $\hat{\rho}(z,t)$ that (in the Schrödinger picture) depends not only on time, but also on the classical variables $z\in\mathfrak{P}$. The states of the classical and quantum subsystems are obtained via
	\begin{align}
		\varrho(z,t) = \Tr[\hat\rho(z,t)] , \label{red-C} \\
		\rho(t) = \! \int \! \hat\rho(z,t) \;\! dz ,   \label{red-Q}
	\end{align}
	respectively. The CQ state is normalized,
	\begin{equation}
		\int \! \Tr[\hat\rho(z,t)] \; dz = 1 , 
		\label{eq:CQ.state.normalization}
	\end{equation}
	with obvious extensions to the expectation values of various operators. A faithful embedding of the full phase space $(q,p)$ with long-time preservation of $q$-$p$ relations is not required for our argumentation. Further mathematical details related to the structure of hybrid systems are provided in Appendix~\ref{app:ss:mathematical.framework}. 
	
	A CP trace-decreasing evolution $\hat\rho(t) = \Lambda[\hat{\rho}(0)]$ is given explicitly via the kernel
	\begin{equation} 
		\hat\rho(z,t) = \! \int \! \Lambda(z|z'\!,t) \;\! \hat\rho(z'\!,0) \;\! dz' .
	\end{equation}
	If this is true for every time $t$, then it is generated by the Gorini--Kossakowski--Sudarshan--Lindblad (GKSL) equation \cite{Gorini.Kossakowski.Sudarshan:JMathPhys:1976,Lindblad:CommunMathPhys:1976,Breuer.Petruccione:book:2007}, which can take several forms when adapted to describe CQ dynamics \cite{Oppenheim:PhysRevX:2023,Oppenheim.etal:NatCommun:2023}. One of these forms is given by 
	\begin{align}
		& \frac{\partial\hat\rho(z,t)}{\partial t} = -i [  H(z,t) , \hat \rho(z,t)] + \lambda^{\mu\nu}(z,t) \Big(\!\!\;   L_\mu\hat\rho(z,t)  L_\nu^\dag \nonumber \\
		& \quad - \tfrac{1}{2}[L_\nu^\dag L_\mu,\hat\rho(z,t)]_+ \!\!\;\Big) + \! \int \! W^{\mu\nu}(z|z'\! ,t) \Big(\!\!\;   L_\mu\hat\rho(z'\!,t)  L_\nu^\dag \nonumber \\
		& \quad - \tfrac{1}{2}[  L_\nu^\dag   L_\mu, \hat{\rho}(z,t)]_+ \!\!\;\Big) \;\! dz' .
		\label{eq:master.equation}
	\end{align}
	Here, the commutator term describes unitary Hamiltonian evolution with classical phase space dependence; the second term represents a Lindblad-type dissipator that is local in phase space; and the integral term introduces nonlocal couplings across classical configurations $z$ and $z'$. The anticommutator of two operators is defined as $[A,B]_+ \defeq AB + BA$, and the kernel
	\begin{equation}
		W^{\mu\nu}(z,t) = \! \int \! W^{\mu\nu}(z'|z,t) \;\! dz'
	\end{equation}
	that multiplies the last term has been absorbed into the integral. Here, $\{L_\mu\}$ is a basis set of operators that we assume to be Hilbert--Schmidt orthogonal (and thus span the space of all trace-class operators). In general, we use the set of operators $\{  L_\mu\}=\{L_0,  L_\alpha\}$ that includes the identity operator on the Hilbert space $\eH$, $ L_0 = \mathbb{I}_\eH$. For finite-dimensional spaces of dimension $d$, it is convenient to choose $L_\alpha$ as the set of generators of $\mathfrak{su}(d)$. Here and in what follows, we use the Einstein convention to sum over repeated indices. For continuous bases, the sum over $\mu\nu$ is replaced by an integral. The coefficients $\lambda^{\mu\nu}$ characterize the strength of various non-Hamiltonian terms.
	
	The probability is conserved,
	\begin{equation}
		\partial_t \! \int \! \Tr[\hat\rho(z,t)] \;\! dz=0,
	\end{equation}
	and the matrices $\lambda^{\mu\nu}(z,t)$, $W^{\mu\nu}(z,t)$, and $\lambda^{\mu\nu}(z,t) + W^{\mu\nu}(z,t)$ satisfy positive semidefiniteness requirements.
	
	This form of the master equation explicitly separates the Hamiltonian and stochastic parts, which allows for a straightforward reduction to purely classical and purely quantum settings. Following Ref.~\cite{Oppenheim:PhysRevX:2023}, we consider only the autonomous case, where there is no time dependence in $\lambda^{\mu\nu}$ and $W^{\mu\nu}$. Then, the dynamics is generated by the Liouvillian superoperator $\cL$, which acts on the operator-valued function $\hat\rho(t,z)$, as follows:
	\begin{equation}
		\cL\hat\rho(t,z) = \partial_t \;\! \hat{\rho}(t,z), \quad \hat\rho(t,z)=e^{\cL t} \;\! \hat{\rho}(0,z).
	\end{equation}
	A useful way to represent this evolution is
	\begin{align}
		\frac{\partial\hat\rho(z,t)}{\partial t} = & - \! \int \! H^{\mu\nu}(z\vert z')  {L}_\mu \hat{\rho}(z'\!,t)   {L}_\nu^\dagger \; dz' \nonumber \\
		& \qquad - \frac{1}{2} H^{\mu\nu}(z) \big[  {L}_\mu^\dagger  {L}_\nu , \hat{\rho}(z,t) \big]_+  , \label{eq:master}
	\end{align}
	where
	\begin{equation}
		H^{\mu\nu}(z) \defeq \! \int \! H^{\mu\nu}(z'\vert z) \;\! dz' ,
	\end{equation} 
	and the amplitudes $H^{\mu\nu}(z|z')$ are constructed from the decomposition of $H$ and $\lambda^{\mu\nu}$, $W^{\mu\nu}$.
	
	It is useful to define $n$-th statistical moments $D^{\mu\nu}_{(n)}$ of the amplitudes $H^{\mu\nu}(z|z')$ via
	\begin{equation}
		D^{\mu\nu}_{(n)i_1\ldots i_n}\!(z') \defeq \frac{1}{n!} \int \! H^{\mu\nu}(z|z')(z_{i_1}-z'_{i_1})\ldots (z_{i_n}-z'_{i_n}) \;\! dz  .
	\end{equation} 
	The master equation then takes the form
	\begin{align}
		& \frac{\partial\hat\rho(z,t)}{\partial t} = \sum_{n=1}^\infty (-1)^n \! \left(\frac{\partial^n}{\partial z_{i_1}\ldots z_{i_n}}\right) \! D^{00}_{(n)i_1\ldots i_n}(z)\hat\rho \nonumber \\
		& \quad -i[  H(z),\hat\rho]+D_{(0)}^{\alpha\beta}(z)  L_\alpha \hat\rho   L_\beta^\dag-\tfrac{1}{2} D_{(0)}^{\alpha\beta}(z) [  L_\beta^\dag  {L}_\alpha , \hat\rho ]_+ \nonumber\\
		& \quad + \! \sum_{\mu,\nu \neq 0} \sum_{n=1}^{\infty} (-1)^n \! \left(\frac{\partial^n}{\partial z_{i_1} \cdots \partial z_{i_n}}\right) \! D^{\mu\nu}_{(n)i_1\ldots i_n}(z) L_\mu \hat{\rho} L_\nu^\dag,
		\label{eq:master.equation.moments}
	\end{align} 
	where $H=\tfrac{i}{2}\big(D_{(0)}^{\mu0}\hat L_\mu-D_{(0)}^{0\mu}\hat L_\mu^\dagger \big)$. The first line of this equation represents the purely classical dynamics that is described by the moments $D^{00}_{(n)}$ generated by the quantum identity components of the evolution. The second line is the GKSL evolution controlled by the classical state $z\in\gP$ and is determined by the zeroth-order moments $D_{(0)}^{\alpha\beta}$. The last line represents the quantum backreaction on the classical system. Particularly important are the first-order moments that can be interpreted as forces exerted by the quantum system Q on the classical system C, and the second-order moments describing the phase space diffusion. 
	
	The three expectation values
	\begin{align}
		\langle D_{(0)} \rangle &= \!\int\! \Tr\big[ D_{(0)}^{\alpha\beta} \;\! L_\alpha \;\! \hat{\rho}(z) \;\! L^\dagger_\beta \big] \;\! dz , \\
		\langle D_{(1)}^{\text{br}} \rangle_i &= \!\int\! \Tr \big[ D_{(1)i}^{\mu\beta} \;\! L_\mu \;\! \hat{\rho}(z) \;\! L^\dagger_\beta \big] \;\! dz ,\\
		\langle D_{(2)}^{\text{br}} \rangle_{ij} &= \!\int\! \Tr \big[ D_{(2)ij}^{\mu\nu} \;\! L_\mu \;\! \hat{\rho}(z) \;\! L^\dagger_\nu \big] \;\! dz ,
	\end{align} 
	are of particular relevance for encoding the backreaction of the quantum system onto the classical background.\ More specifically, the necessary condition for CQ dynamics with quantum backreaction is $\langle D_{(0)} \rangle > 0$. In addition, the moments have to satisfy the diffusion-decoherence relation \cite{Oppenheim.etal:NatCommun:2023} 
	\begin{equation}
		2 \langle D_{(2)}^{\text{br}} \rangle_{ij} \langle D_{(0)}\rangle \geqslant \langle D_{(1)}^{\text{br}} \rangle_i \langle D_{(1)}^{\text{br}} \rangle_j 
		\label{eq:diffusion.decoherence.relation}
	\end{equation} 
	for all CQ states $\hat\rho$ and indices $i,j$.

	\section[Fundamental issues arising \\ from completely positive dynamics]{Fundamental issues arising \newline from completely positive dynamics} \label{sec:CP.fundamental.issues}
	The consistency requirements imposed on CQ hybrid systems are summarized in Appendix~\ref{app:ss:consistency.requirements}. There are two distinct types: the majority deal with formal mathematical or structural properties, such as positivity of classical and quantum probabilities, implementation of canonical transformations, and the classical limit of the quantum and hybrid CQ systems (requirements I--IV, VI, and VII). The remainder (requirement V) concerns compatibility with fundamental physical constraints, such as conservation laws, relativistic causality, and the second law of thermodynamics \cite{Layton.Miller:2504.10587}. 
	
	For a hybrid model that is used as a computational tool, where both subsystems are in fact quantum but one is treated classically for efficiency, violation of these requirements at some point in the evolution simply sets the timescale over which the scheme remains valid. 
	
	In contrast, the same conditions are interpreted much more stringently for hybrid models purporting to be a fundamental theory: there is no `expiration date' on the formal and structural requirements, and this is precisely where unitary models without additional dissipation and diffusion terms generally fail \cite{Terno:chapter:2025}. The scheme proposed in Ref.~\cite{Oppenheim:PhysRevX:2023} on the other hand satisfies them by construction.
	
	From an empirical point of view, the relevant consistency requirement is compatibility with current experimental data. Hybrid dynamics may violate the conservation laws that, in our present effective theories, follow from underlying symmetry principles, but any such violations must lie within existing experimental bounds in the regimes where the hybrid model is intended to apply (and, ideally, do so without requiring fine tuning). In a CP hybrid model applied to gravity interacting with quantum matter, information loss is a predetermined outcome. However, it also leads to a violation of energy conservation in a setting where this concept is perfectly well-defined. Unruh and Wald argued that such violations are unobservable for astrophysical-scale black holes \cite{Unruh.Wald:PhysRevD:1995}. 	
	
	We consider a classical register $z$ whose statistics are represented by a quantum state diagonal in the $\lbrace \vert z \rangle \}$ basis, and we assume this $z$-diagonality is preserved by the dynamics (e.g., a pointer basis stabilized by efficient decoherence that maintains it despite the interactions). Such a quantum system cannot mediate entanglement, and thus the empirical detection of gravity-mediated entanglement would falsify this hybrid scheme \cite{Marletto.Vedral:RevModPhys:2025}.  
	
	A CQ hybrid theory should also be internally consistent: positivity requirements ultimately translate into relations such as Eq.~\eqref{eq:diffusion.decoherence.relation}, which are responsible for the trade-off between dissipation and diffusion described in Ref.~\cite{Oppenheim.etal:NatCommun:2023}. However, the coefficients $D_{(n)}$ for $n=0,1$ are determined by the standard equations of general relativity (for example, Hamiltonian evolution in pure gravity, such as the propagation of gravitational waves in vacuum) and the unitary evolution of quantum fields on a fixed background. Therefore, the consistency of such relations with the underlying theory should be analyzed separately.
	
	The violation of energy conservation is not the only potentially problematic consequence resulting from the CP description of a fundamental theory. Since the evolution under $e^{\cL t}$ is no longer unitary, the possible trajectories of $\hat\rho(t)$ and the asymptotic states of the system for $t\to\infty$ are determined by possibly degenerate eigenvalues of $\cL$. 
	
		\begin{table*}[htbp!] 
		\centering
		\resizebox{\textwidth}{!}{
			\begin{tabular}{>{\raggedright\arraybackslash}m{0.17\linewidth} 
					@{\hskip 0.01\linewidth} >{\centering\arraybackslash}m{0.1875\linewidth} 
					@{\hskip 0.01\linewidth} >{\centering\arraybackslash}m{0.1875\linewidth} 
					@{\hskip 0.03\linewidth} >{\centering\arraybackslash}m{0.1875\linewidth} 
					@{\hskip 0.03\linewidth} >{\centering\arraybackslash}m{0.1875\linewidth}} 
				\toprule \toprule
				\textbf{Theory} & \textbf{Symmetry of EOM} & \textbf{Local Conservation} & \textbf{Global Conservation} & \textbf{Noether's Theorem} \\ 
				\midrule \\[-0.5mm]
				Classical Mechanics & \cmark & n/a & \cmark & \cmark \\[4mm]
				\multirow{2}{=}{Classical Field Theory} 
				& \multirow{2}{*}{\cmark} 
				& \multirow{2}{*}{\cmark} 
				& \multirow{2}{*}{\cmark} 
				& \multirow{2}{*}{\cmark} 
				\\[7mm]
				\multirow{2}{=}{Quantum Mechanics (Closed Systems)} 
				& \multirow{2}{*}{\cmark} 
				& \multirow{2}{*}{n/a} 
				& \multirow{2}{*}{\cmark} 
				& \multirow{2}{*}{\cmark} 
				\\[7mm]
				\multirow{2}{=}{Open Quantum Systems (Lindblad)} 
				& \multirow{2}{*}{\cmark} 
				& \multirow[c]{2}{*}{\makecell[c]{\faExclamationTriangle \\ (dissipation, decoherence)}}
				& \multirow[c]{2}{*}{\makecell[c]{\xmark \\ (only full system + bath)}}
				& \multirow{2}{*}{\xmark} 
				\\[7mm]
				\multirow{2}{=}{General Relativity} 
				& \multirow{2}{*}{\cmark} 
				& \multirow{2}{*}{\cmark} 
				& \multirow[c]{2}{*}{\makecell[c]{\faExclamationTriangle \\ (not well-defined)}}
				& \multirow{2}{*}{\xmark} 
				\\[7mm]
				\multirow{2}{=}{Quantum Field Theory (Flat Spacetime)} 
				& \multirow{2}{*}{\cmark} 
				& \multirow{2}{*}{\cmark} 
				& \multirow[c]{2}{*}{\makecell[c]{\cmark \\ (if no quantum anomaly)}} 
				& \multirow[c]{2}{*}{\makecell[c]{\cmark \\ (if no quantum anomaly)}}
				\\[7mm]
				\multirow{2}{=}{Quantum Field Theory (Curved Spacetime)} 
				& \multirow[c]{2}{*}{\makecell[c]{\faExclamationTriangle \\ (depends on background)}}
				& \multirow[c]{2}{*}{\makecell[c]{\faExclamationTriangle \\ (anomalies, backreaction)}}
				& \multirow[c]{2}{*}{\makecell[c]{\xmark \\ (no global Killing vector)}}
				& \multirow[c]{2}{*}{\makecell[c]{\faExclamationTriangle \\ (trace anomaly)}}
				\\[7mm]
				\multirow{2}{=}{Classical-Quantum Hybrid Models} 
				& \multirow{2}{*}{\cmark}
				& \multirow{2}{*}{\centering\makecell{\xmark \\ (no continuity equation)}} 
				& \multirow{2}{*}{\xmark}
				& \multirow{2}{*}{\centering\makecell{\xmark \\ (no global action)}} 
				\\[5mm]
				\bottomrule \bottomrule
			\end{tabular}
		}
		\caption{Symmetries, conservation laws, and Noether's theorem across different theoretical frameworks.}
		\label{tab:overview}
		\vspace*{-3mm}
	\end{table*}
	
	In terms of their physical significance, one distinguishes four types of eigenvalues of $\cL$ \cite{Baumgartner.Narnhofer:JPhysA:2008,Albert.Jiang:PhysRevA:2014}: zero eigenvalues correspond to stationary states $\hat\rho_\infty$ (where we have suppressed the degeneracy labels for the moment). Purely imaginary eigenvalues correspond to circular paths, meaning that there are starting configurations that will not approach an equilibrium state in the long-time limit. However, they appear only under very restrictive conditions that are typically not realized in physically relevant situations \cite{Baumgartner.Narnhofer:JPhysA:2008,Nigro:JStatMechTheoryExp:2019}. Lastly, real negative eigenvalues directly lead to zero, while complex pairs of eigenvalues with negative real part spiral to zero.  
	
	The set of asymptotic states $\hat\rho_\infty$ forms an asymptotic subspace $\mathrm{As}(\eH)$ of the state space \cite{Albert.Jiang:PhysRevA:2014,Albert.etal:PhysRevX:2016}. It attracts all initial states $\hat\rho(t=0)$ and is free from decoherence. Any remaining time evolution within $\mathrm{As}(\eH)$ is unitary \cite{Albert.Jiang:PhysRevA:2014}.
	
	For a large number of systems, particularly finite-dimensional ones, the steady state is unique. In any case, $\mathrm{As}(\eH)$ is generically a proper subspace of the state space, and a generic initial state $\hat\rho(t=0)$ asymptotically evolves into a block-diagonal form. Several results describe how particular initial states are mapped into such blocks, also known as basins \cite{Baumgartner.Narnhofer:JPhysA:2008}.
	
	This has important consequences for conservation laws \cite{Albert.Jiang:PhysRevA:2014,Albert.etal:PhysRevX:2016,Ulcakar.Lenarcic:PhysRevLett:2024}: In a quantum system that evolves unitarily, a time-independent self-adjoint operator $A$ represents a conserved observable if and only if it commutes with the system's Hamiltonian $H$. Analogously, for a classical phase-space function, the conservation is equivalent to the vanishing of the Poisson bracket with the Hamiltonian. According to Noether's theorem \cite{Noether:1918}, continuous symmetries of the Lagrangian result in conserved quantities. In Hamiltonian systems, these conserved quantities typically serve as generators of the associated symmetry transformations. Table~\ref{tab:overview} provides an overview of local and global conservation laws as well as the applicability of Noether's theorem in different theoretical frameworks.
	
	For systems with CP dynamics, the conservation of such an observable naturally means \cite{Albert.Jiang:PhysRevA:2014}
	\begin{equation} 
		dA/dt=0 \quad \Longleftrightarrow \quad \cL^\dag(A)=0 .
	\end{equation}
	As a result, the relation between symmetries and conserved quantities established by Noether's theorem is broken. On the one hand, there may exist conserved quantities that do not commute with every term in $\cL$. Consequently, not all unitary transformations that such observables generate are symmetries of the system. On the other hand, a symmetry generator does not have to be a conserved quantity. This motivates asking whether the completely positive classical-quantum (CPCQ) dynamics is fundamental or instead a reduced description of a larger quantum theory.
	
	Apart from $t\to\infty$ asymptotics, other timescales may be important \cite{Macieszczak.etal:PhysRevLett:2016}. Due to the splitting of the spectrum of $\cL$, a partial relaxation into long-lived metastable states is possible. These are distinct from the asymptotic stationary states $\rho_\infty$. Metastability manifests itself as a mid- or long-time regime when the system appears stationary, before eventually approaching $\mathrm{As}(\eH)$. For specific models, it is possible to obtain a low-dimensional approximation of the dynamics in terms of the motion in a manifold of metastable states. The separation between the low-lying eigenvalues and the rest of the spectrum occurs at some $\lambda_{m+1}$, $\vert \mathrm{Re}[\lambda_{m+1}] \vert \gg \vert \mathrm{Re}[\lambda_{m}] \vert$. The upper limit on the duration of the metastable interval is then determined by $1/{\vert \mathrm{Re}[\lambda_m]\vert}$.
	
	\section[Embedding classical-quantum gravity \\ into a fully quantum theory]{Embedding classical-quantum gravity \newline into a fully quantum theory}
	\label{sec:embedding.CQgravity.into.Qtheory}	
	We now show that, starting from the Kraus representation \eqref{eq:CQ.state.Kraus.representation}, the embedding for CPCQ dynamics can be made explicit for its discrete counterpart \eqref{eq:discrete.CQ.state.Kraus.representation}. More precisely, if the classical label $z$ is encoded in an orthonormal basis of an auxiliary system $Z$, then the CQ evolution is recovered as the $Z$-diagonal sector of a completely positive trace-preserving (CPTP) map on $QZ$, together with an operational restriction to the commutative algebra $\mathcal{A}_Z \defeq \mathrm{span}\{\Pi_z\}_z$, where $\Pi_z \defeq \vert z\rangle_Z \langle z\vert_Z$.\footnote{For a continuum phase space, the same idea requires either a generalized-basis/direct-integral treatment or an explicit coarse graining to a discrete classical register.}\ This is not a canonical quantization of the classical phase-space variables; rather, it realizes the hybrid theory as an operationally restricted sector of a larger quantum channel. Equivalently, the same construction may be viewed as a reduced/open description obtained from a unitary Stinespring completion. Related embeddings or realizations of CPCQ dynamics have appeared previously in abstract diagonal form \cite{Blanchard.Jadczyk:PhysLettA:1993,Diosi:PhysRevA:2023} and in bosonic phase-space realizations of the CQ limit \cite{Layton.Oppenheim:PRXQuantum:2024}.
	
	In what follows, we denote the CQ state by $\hat{\rho}_{Q}(z,t)\in\mathcal{T}_1(\eH_Q)$, where $\mathcal{T}_1(\eH_Q)$ denotes the trace-class operators on $\eH_Q$, and use subscripts to indicate operator support: $K_{\alpha Q},\,L_{\beta Q},\,\mathbb{I}_Q$ act on $Q$, while $\Xi_{\alpha}(t)_{QZ},\,\mathbb{I}_{QZ}$ act on $QZ$.
	
	In the formulation of Ref.~\cite{Oppenheim:PhysRevX:2023}, a generic CQ state is described by [Eq.~(28) of Ref.~\citenum{Oppenheim:PhysRevX:2023}]
	\begin{equation}
		\hat{\rho}_{Q}(z,t) = \!\int \sum\limits_{\alpha } K_{\alpha Q}(z|z' \!,t) \;\! \hat{\rho}_{Q}(z' \!,0) \;\! K_{\alpha Q}(z|z' \!,t)^{\dagger} \;\! dz' ,
		\label{eq:CQ.state.Kraus.representation}
	\end{equation}
	with the Kraus operators satisfying [Eq.~(29) of Ref.~\citenum{Oppenheim:PhysRevX:2023}]
	\begin{equation}
		\sum\limits_{\alpha} \int K_{\alpha Q}(z|z'\!,t)^\dagger \;\! K_{\alpha Q}(z|z''\!,t) \;\! dz = \delta (z' \!\!-\! z'') \;\! \mathbb{I}_{Q} .
	\end{equation}
	The CQ state can also be expressed as [Eq.~(34) of Ref.~\citenum{Oppenheim:PhysRevX:2023}]
	\begin{equation}
		\hat{\rho}_{Q}(z,t) = \!\int \sum\limits_{\beta \beta'} \Lambda _{\beta \beta'}(z|z'\!, t) \left( L_{\beta Q} \;\! \hat{\rho}_{Q}(z' \!,0) \;\! L_{\beta' Q}^{\dagger} \right) dz' . 
	\end{equation}
	Let us consider the discrete versions of these equations:
	\begin{equation}
		\hat{\rho}_Q(z,t) = \sum\limits_{z'} \sum\limits_{\alpha} K_{\alpha Q}(z|z'\!,t) \;\! \hat{\rho}_Q(z'\!,0) \;\! K_{\alpha Q}(z|z'\!,t)^{\dagger},
		\label{eq:discrete.CQ.state.Kraus.representation}
	\end{equation}
	\begin{equation}
		\sum\limits_{\alpha,z} K_{\alpha Q}(z|z'\!,t)^\dagger \;\! K_{\alpha Q}(z|z''\!,t) = \delta_{z'\! z''} \;\! \mathbb{I}_{Q},
	\end{equation}
	\begin{equation}
		K_{\alpha Q}(z|z'\!,t) = \sum_{\beta} G_{\alpha\beta}(z|z'\!,t) \;\! L_{\beta Q},
	\end{equation}
	\begin{equation}
		\Lambda _{\beta \beta'}(z|z'\!,t) = \sum\limits_{\alpha} G_{\alpha \beta}(z|z'\!,t) \;\! G_{\alpha \beta'}(z|z'\!,t)^{*},
	\end{equation}
	\begin{equation}
		\hat{\rho}_{Q}(z,t) = \sum\limits_{\beta \beta'} \sum\limits_{z'} \Lambda _{\beta \beta'}(z|z'\!,t) \left( L_{\beta Q} \;\! \hat{\rho}_{Q}(z'\!,0) \;\! L_{\beta' Q}^{\dagger} \right) ,
	\end{equation}
	where $G_{\alpha\beta}(z|z'\!,t)$ and $G_{\alpha\beta}(z|z'\!,t)^{*}$ denote the scalar expansion coefficients of the Kraus operators in a fixed Hilbert--Schmidt orthonormal operator basis $\{L_{\beta Q}\}$ on $\eH_{Q}$ and their complex conjugates, respectively.
	
	We now embed the CQ dynamics into a CPTP evolution on $QZ$ and show that the classicality of $z$ follows from a measurement restriction on $Z$ to the commutative algebra $\mathcal{A}_Z \defeq \mathrm{span}\{\Pi_z\}_z$, with $\Pi_z \defeq \vert z\rangle\langle z\vert \equiv \vert z\rangle_Z \langle z\vert_Z$.	We define time-dependent Kraus operators acting on the composite $QZ$ Hilbert space explicitly by
	\begin{equation}
		\Xi_{\alpha}(t)_{QZ} \defeq \sum_{z,z'} K_{\alpha Q}(z|z'\!,t) \;\! \otimes \;\! \vert z\rangle \langle z'\vert .
	\end{equation}
	While they are not unique, such operators always exist. The Kraus completeness on $QZ$ is then equivalent to the CQ completeness:
	\begin{align}
		\begin{aligned}
			\sum\limits_{\alpha} \Xi_{\alpha}(t)_{QZ}^{\dagger} \; \Xi_{\alpha}(t)_{QZ} &= \mathbb{I}_{QZ}. \\
			\Longleftrightarrow\;\;\!
			\sum_{\alpha,z} K_{\alpha Q}(z|z'\!,t)^{\dagger} \;\! K_{\alpha Q}(z|z''\!,t) &= \delta_{z'\! z''} \;\! \mathbb{I}_Q .
		\end{aligned}
	\end{align}
	We can define a time-dependent quantum channel of the $QZ$ system as
	\begin{equation}
		\Gamma_{\! t} \!\left[\rho_{QZ}\right] \defeq \sum\limits_{\alpha} \Xi_{\alpha}(t)_{QZ} \;\! \rho_{QZ} \;\! \Xi_{\alpha}(t)_{QZ}^{\dagger}.
	\end{equation}
	The Stinespring representation \cite{Stinespring:ProcAmerMathSoc:1955} of $\Gamma_{\! t}$ is given by 
	\begin{equation}
		\Gamma_t[\rho_{QZ}] = \Tr_{E} \!\Big[U_{QZE}(t) \;\! \big(\rho_{QZ} \!\otimes\! \vert 0\rangle_E \langle 0\vert_E \big) \;\! U_{QZE}(t)^\dagger\Big],
		\label{eq:Stinespring.rep.embedding}
	\end{equation}
	where $U_{QZE}(t)$ is a unitary time evolution operator, $E \equiv \bar{Q}\bar{Z}$ denotes the ancilla system, and $\vert 0\rangle_{E}$ its initial state. The initial state of the $QZ$ system is described by   
	\begin{equation}
		\rho_{QZ}(0) = \sum\limits_{z} \hat{\rho}_{Q}(z,0) \;\! \otimes \;\! \Pi_z.
	\end{equation}	
	Its state at time $t$ is given by
	\begin{equation}
		\rho_{QZ}(t) \defeq \sum_{z} \big( \mathbb{I}_{Q} \otimes \Pi_z \big) \;\! \Gamma_{\! t} \!\left[ \rho_{QZ}(0)\right] \;\! \big( \mathbb{I}_{Q} \otimes \Pi_z \big),
		\label{eq:rhoQZ.embedding}
	\end{equation}
	which can be rewritten as
	\begin{equation}
		\rho_{QZ}(t) = \sum_{z} \langle z \vert \;\! \Gamma_{\! t}\!\left[\rho_{QZ}(0)\right] \vert z\rangle \;\! \otimes \;\! \Pi_z.
	\end{equation}
	Here, the first term can be computed as follows:
	\begin{widetext}
		\begin{align}
			\begin{aligned}
				\langle z \vert \;\! \Gamma_{\!t}[\rho_{QZ}(0)] \vert z \rangle &= \sum_{\alpha} \big\langle z \big\vert \;\! \Xi_{\alpha}(t)_{QZ} \;\! \rho_{QZ}(0) \;\! \Xi_{\alpha}(t)_{QZ}^{\dagger} \;\! \big\vert z \big\rangle = \sum_{\alpha,z',z''} K_{\alpha Q}(z|z'\!,t) \;\! \big\langle z' \big\vert \;\! \rho_{QZ}(0) \;\! \big\vert z'' \big\rangle \;\! K_{\alpha Q}(z|z''\!,t)^{\dagger} \\
				&= \sum_{\alpha,z'} K_{\alpha Q}(z|z'\!,t) \;\! \hat{\rho}_Q(z'\!,0) \;\! K_{\alpha Q}(z|z'\!,t)^{\dagger} = \hat{\rho}_Q(z,t).
			\end{aligned}
		\end{align}
	\end{widetext}
	Thus, we obtain
	\begin{equation}
		\langle z \vert \;\! \Gamma_{\! t} \!\left[ \rho_{QZ}(0)\right] \vert z\rangle = \hat{\rho}_{Q}(z,t) ,
	\end{equation}
	where $\hat{\rho}_{Q}(z,t)$ is the CQ state described in Eq.~\eqref{eq:discrete.CQ.state.Kraus.representation}. This implies that the relation  
	\begin{equation}
		\rho_{QZ}(t) = \sum\limits_{z} \hat{\rho}_{Q}(z,t) \;\! \otimes \;\! \Pi_z
	\end{equation}
	holds at all times $t$.
	
	To make the $Z$-restriction operational, note that the $Z$-pinching map
	\begin{equation}
		\mathcal{P}_Z[X] \defeq \sum_{z}(\mathbb{I}_Q \otimes \Pi_z) \;\! X \;\! (\mathbb{I}_Q \otimes \Pi_z)
	\end{equation}
	produces Eq.~\eqref{eq:rhoQZ.embedding} directly:
	\begin{equation}
		\rho_{QZ}(t) = \mathcal{P}_Z \big[\Gamma_{\! t}[\rho_{QZ}(0)]\big] =\sum_{z} \hat{\rho}_{Q}(z,t) \;\! \otimes \;\! \Pi_z .
	\end{equation}
	For any allowed observable $A_Q \!\otimes\! A_Z$ with $A_Z \in \mathcal{A}_Z$,
	\begin{equation}
		\Tr \big[(A_Q \!\otimes\! A_Z) \;\! X \big] = \Tr \big[(A_Q \!\otimes\! A_Z) \;\! \mathcal{P}_Z[X] \big].
	\end{equation}
	Hence the restriction renders $z$ operationally classical, since $A_Z=\sum_z a_z \Pi_z$ and $\Pi_z\Pi_{z'}=\delta_{zz'}\Pi_z$ eliminate off-diagonal $Z$ blocks.
	
	The Lüders instrument \cite{Lueders:AnnPhys:1951} for the projectors $\{\Pi_z\}$ is
	\begin{equation}
		\mathcal{I}_z[X] \defeq (\mathbb{I}_Q \!\otimes\! \Pi_z) \;\! X \;\! (\mathbb{I}_Q \!\otimes\! \Pi_z), 
		\quad
		\sum_{z} \mathcal{I}_z = \mathcal{P}_Z .
	\end{equation}
	Thus Eq.~\eqref{eq:rhoQZ.embedding} is the \emph{nonselective} post-measurement update on $Z$ applied after $\Gamma_{\! t}$, whereas conditioning on outcome $z$ yields the \emph{selective} state 
	\begin{align}
		\begin{aligned}
			\rho_Q^{(t|z)} &= \frac{\langle z \vert \;\! \Gamma_{\! t}[\rho_{QZ}(0)] \;\! \vert z \rangle}{p_t(z)} \\
			&= \frac{\Tr_{Z}\!\big[(\mathbb{I}_Q \!\otimes\! \Pi_z) \;\! \Gamma_{\! t}[\rho_{QZ}(0)] \;\! (\mathbb{I}_Q \!\otimes\! \Pi_z)\big]}{p_t(z)} \\
			&= \frac{\hat{\rho}_Q(z,t)}{p_t(z)} ,
		\end{aligned}
	\end{align}
	with $p_t(z) = \Tr[\hat{\rho}_Q(z,t)]$. Therefore, the CPCQ dynamics can be embedded into a fully quantum theory on an enlarged Hilbert space.
	
	Finally, we emphasize that this embedding is not a standard canonical quantization, since the classical phase-space coordinate $z$ collects both the spatial metric $h$ and its canonical conjugate momentum $\pi$. In the construction above, these variables are encoded as mutually commuting observables on the auxiliary Hilbert space $\eH_Z$, such that $[h,\pi]=0$ in this representation. The system $Z$ thus provides a commutative pointer algebra with preferred basis $\{\vert z\rangle_Z\}_z$, rather than a canonical phase-space representation with nontrivial commutation relations. Nevertheless, despite this noncanonical choice, the total CQ system is still realized as a fully quantum system $QZ$ (and its Stinespring dilation $QZE$) governed by the CPTP map $\Gamma_{\! t}$.

	\section[Angular-momentum nonconservation in a \\ rotationally invariant classical-quantum system]{Angular-momentum nonconservation in a rotationally invariant classical-quantum system} \label{sec:toy.model}
	Noether's theorem \cite{Noether:1918} does not apply in CQ frameworks due to the lack of a unified action principle that treats both classical and quantum components on an equal footing. While this does not \emph{a priori} preclude the possibility that symmetry generators are conserved, it implies that their conservation cannot be derived using traditional (i.e., based on a global action-based symmetry principle) variational and/or symplectic methods. Having established above that CPCQ dynamics admits a fully quantum realization on an enlarged Hilbert space, we now turn to a complementary issue: the fate of conservation laws in the reduced hybrid description. To illustrate this concretely, we study a simple toy model in which angular momentum is not conserved under the dynamical evolution prescribed by the master equation \eqref{eq:master.equation}.
	
	Consider a qubit interacting with a classical particle in 3+1 dimensions.\ According to Eq.~\eqref{eq:master.equation}, the dynamics of this coupled system is described by 
	\begin{widetext}
		\begin{equation}
			\frac{\partial}{\partial t} \hat\rho(\vec{q},\vec{p},t) = - \kappa \;\! \hat{\rho}(\vec{q},\vec{p},t) + \frac{\kappa}{4} \;\! \rho_{\mathrm{f}}(\vert \vec{q} \vert, \vert \vec{p} \vert, \vec{q} \cdot \vec{p}) \int\! \Big( \hat{\rho}(\vec{q}^{\,\prime\!\!},\vec{p}^{\,\prime\!\!},t) + \sum_a \hat{\sigma}_a \;\! \hat{\rho}(\vec{q}^{\,\prime\!\!},\vec{p}^{\,\prime\!\!},t) \;\! \hat{\sigma}_a \Big) \;\! d^3 \! q' \;\! d^3 \! p' ,
		\end{equation}
		where $a \in \lbrace x,y,z \rbrace$, $\kappa$ corresponds to the decoherence rate, and $\rho_{\mathrm{f}}(\vert \vec{q} \vert, \vert \vec{p} \vert, \vec{q} \cdot \vec{p})$ specifies the rotationally invariant phase-space distribution of the stationary CQ state, $\hat{\rho}_{\infty}(\vec{q},\vec{p}) = \tfrac{\mathbb{I}}{2}\rho_{\mathrm{f}}(\vert \vec{q} \vert, \vert \vec{p} \vert, \vec{q} \cdot \vec{p})$.\footnote{Note that $\lbrace \vert \vec{q} \vert, \vert \vec{p} \vert, \vec{q} \cdot \vec{p} \rbrace$ form a minimal rotationally invariant set of scalars that fully characterizes the dependence of $\rho_{\mathrm{f}}$ on classical variables.}\ The solution of this equation is given by
		\begin{align}
			\hat\rho(\vec{q},\vec{p},t) &= e^{-\kappa t} \hat{\rho}(\vec{q},\vec{p},0) + \tfrac{1}{4} \! \left(1-e^{-\kappa t}\right) \rho_{\mathrm{f}}(\vert \vec{q} \vert, \vert \vec{p} \vert, \vec{q} \cdot \vec{p}) \nonumber \\
			& \quad \int \! \Big( \hat{\rho}(\vec{q}^{\,\prime\!\!},\vec{p}^{\,\prime\!\!},0) + \hat{\sigma}_x \;\! \hat\rho(\vec{q}^{\,\prime\!\!},\vec{p}^{\,\prime\!\!},0) \;\! \hat{\sigma}_x + \hat{\sigma}_y \;\! \hat\rho(\vec{q}^{\,\prime\!\!},\vec{p}^{\,\prime\!\!},0) \;\! \hat{\sigma}_y + \hat{\sigma}_z \;\! \hat\rho(\vec{q}^{\,\prime\!\!},\vec{p}^{\,\prime\!\!},0) \;\! \hat{\sigma}_z \Big) \;\! d^3 \! q' \;\! d^3 \! p' .
			\label{eq:toy.model.solution.full}
		\end{align}
	\end{widetext}
	Using the normalization \eqref{eq:CQ.state.normalization} and the identity
	\begin{equation}
		2 \Tr[\hat\rho] \;\! \mathbb{I} = \hat{\rho} + \hat{\sigma}_x \hat{\rho} \hat{\sigma}_x + \hat{\sigma}_y \hat{\rho} \hat{\sigma}_y + \hat{\sigma}_z \hat{\rho} \hat{\sigma}_z ,
	\end{equation}
	the solution \eqref{eq:toy.model.solution.full} simplifies to 
	\begin{equation}
		\hat{\rho}(\vec{q},\vec{p},t) = e^{-\kappa t} \hat{\rho}(\vec{q},\vec{p},0) + \tfrac{\mathbb{I}}{2} \! \left(1-e^{-\kappa t}\right) \rho_{\mathrm{f}}(\vert \vec{q} \vert, \vert \vec{p} \vert, \vec{q} \cdot \vec{p}) .
	\label{eq:toy.model.solution.simplified}
	\end{equation}
	Let us now suppose that the initial state is given by
	\begin{equation}
		\hat{\rho}(\vec{q},\vec{p},0) = \delta^3 (\vec{q}-\vec{q}_0) \;\! \delta^3 (\vec{p}-\vec{p}_0) \;\! \hat\rho_i ,
	\end{equation}
	where $\hat{\rho}_i$ is a constant density matrix, the initial orbital angular momentum is given by $\vec{L}_0 = \vec{q}_0 \times \vec{p}_0$, and the initial spin operator expectation values are
	\begin{equation}
		\tfrac{\hbar}{2} \;\! \big\langle \sigma_a(0) \big\rangle = \!\int\! \Tr[\hat{\sigma}_a \hat{\rho}(\vec{q},\vec{p},0)] \;\! d^3 \! q \;\! d^3 \! p = \tfrac{\hbar}{2} \Tr[\hat{\sigma}_a \;\! \hat{\rho}_i] .
	\end{equation}
	The total angular momentum components at $t=0$ are given by
	\begin{equation}
		J_a(0) = \big(\vec{q}_0 \times \vec{p}_0\big)_a + \tfrac{\hbar}{2} \Tr[\hat{\sigma}_a \;\! \hat{\rho}_i] . 
	\end{equation}
	The spin expectation values evolve as
	\begin{equation}
		\tfrac{\hbar}{2} \;\! \big\langle \sigma_a(t) \big\rangle = \!\int\! \Tr[\hat{\sigma}_a \;\! \hat{\rho}(\vec{q},\vec{p},t)] \;\! d^3 q \;\! d^3 p = \tfrac{\hbar}{2} \;\! e^{-\kappa t} \Tr[\hat\sigma_a \;\! \hat{\rho}_i] .
	\end{equation}
	Let us assume that the final state has zero orbital angular momentum, $\vec{L}_\mathrm{f}=\vec{0}$, which can be realized by taking
	\begin{equation}
		\rho_{\mathrm{f}}(\vec{q},\vec{p}) = \delta^3(\vec{q}) \;\! \delta^3(\vec{p}) .
	\end{equation}
	Then
	\begin{align}
		\frac{d}{dt} \langle L_a \rangle = - \kappa \;\! \langle L_a \rangle , \quad \frac{d}{dt} \langle \sigma_a \rangle = - \kappa \;\! \langle \sigma_a \rangle ,
	\end{align}
	and thus
	\begin{align}
		\langle J_a(t) \rangle = e^{- \kappa t} \langle J_a(0) \rangle .
	\end{align}
	However, in this case the expectation value of the total angular momentum vanishes as $t\to\infty$, i.e., $\langle J_a(\infty) \rangle = 0$, and using Eq.~\eqref{eq:toy.model.solution.simplified}, we obtain
	\begin{align}
		\hat\rho(\vec{q},\vec{p},t) &= e^{-\kappa t} \delta^3 (\vec{q}-\vec{q}_0) \;\! \delta^3(\vec{p}-\vec{p}_0) \;\! \hat{\rho}_i \nonumber \\
		& \quad + \tfrac{\mathbb{I}}{2} (1-e^{-\kappa t}) \;\! \delta^3(\vec{q}) \;\! \delta^3(\vec{p}) .
	\end{align}
	Despite the rotational symmetry of the equation of motion, this model breaks the conservation of total angular momentum.

	\section{Discussion and conclusions} \label{sec:discussion.conclusions}
	As shown in Section~\ref{sec:embedding.CQgravity.into.Qtheory}, the hybrid description admits a fully quantum realization on an enlarged Hilbert space, but two important consistency questions remain: First, requirements such as Eq.~\eqref{eq:diffusion.decoherence.relation} must be compatible with the coefficients $D_{(n)i_1\ldots i_n}$ determined from general relativity and quantum field theory on curved backgrounds. Second, compatibility with conservation laws presents an additional challenge. While angular momentum is not conserved in our rotationally invariant toy model (Section~\ref{sec:toy.model}), and similar CQ toy models as well as simple semiclassical backreaction lead to nonconservation of energy \cite{Oppenheim.etal:Quantum:2023,Unruh.Wald:PhysRevD:1995}, it is not yet known whether the fully relativistic scheme proposed in Ref.~\cite{Oppenheim:PhysRevX:2023} exhibits the same properties.
	
	If such an incompatibility is indeed present, it is possible that the relevant violation timescale significantly exceeds our lifetime, or even that of our universe. Therefore, even if a hybrid model predicts that conservation laws may be violated in gravity-matter interactions, such violations may not be detectable by quasilocal observers. 
	
	Physically, the fact that a fully closed CQ system can violate conservation laws is explained as follows: In the CP hybrid formalism, the same GKSL coefficients that determine the decoherence rate $\kappa$ also enter the three moments appearing in the diffusion-decoherence inequality \eqref{eq:diffusion.decoherence.relation}. Any nonzero backreaction $\langle D_{(0)} \rangle \!>\! 0$ therefore requires a minimum dissipation budget, which implies a lower bound on $\kappa$ for a fixed backreaction scale. Experimentally, if no relaxation is observed over a timescale $T$, i.e., if $\kappa \lesssim 1/T$, Eq.~\eqref{eq:diffusion.decoherence.relation} directly constrains the allowed regions of $\langle D_{(0)} \rangle$, $\langle D^\text{br}_{(1)} \rangle$, and $\langle D^\text{br}_{(2)} \rangle$. Bounds on the rate $\kappa$ therefore translate directly into bounds on the CQ backreaction and diffusion.
	
	On the other hand, exact conservation laws are restored in a suitably taken limit of vanishing dissipative terms \cite{Ulcakar.Lenarcic:PhysRevLett:2024}. Gravity and matter never really decouple, even though for the most part this interaction is relatively weak. This makes the very early universe, where the effective gravitational interaction was strong, a particularly interesting avenue to explore.
	
	A complementary analysis \cite{Carney.Matsumura:ClassQuantumGrav:2024} showed that a tree-level scattering theory in which classical fields mediate interactions between quantum systems can be formulated in an appropriate limit of the CQ model proposed by Oppenheim and collaborators \cite{Oppenheim:PhysRevX:2023,Oppenheim.etal:NatCommun:2023}. While the scattering probabilities are Lorentz invariant in this framework and total probability is conserved, they violate energy-momentum conservation. This provides further evidence that invariance under spacetime symmetries in CQ models does not, in general, entail conservation of the corresponding Noether charges and highlights a fundamental structural limitation of hybrid approaches based on CP dynamics.
	
	Furthermore, Ref.~\cite{Feng.Marletto.Vedral:2311.08971} argued that for any CQ dynamics that satisfies a global conservation law, the classical subsystem cannot induce changes in the local observables of the quantum subsystem, provided that the evolution can be described either by a Hamiltonian formalism or by a sequence of infinitesimal time steps. This is consistent with the results of our analysis, which show that CQ hybrid models based on CP maps generally violate conservation laws.
	
	Section~\ref{sec:embedding.CQgravity.into.Qtheory} shows that CQ dynamics can be realized as the $Z$-diagonal sector of a CPTP map on the enlarged system $QZ$, or equivalently as a reduced/open description obtained from a unitary Stinespring completion together with an operational restriction to the commutative algebra generated by $\{\Pi_z\}$. The construction is explicitly noncanonical: the coordinate $z$ is represented on $Z$ by mutually commuting observables rather than by canonically conjugate operators.
	
	This perspective clarifies what is genuinely quantum in the CQ framework. The stochastic term is a CPTP noise channel admitting a quantum dilation to an auxiliary environment whose statistics are fixed independently of the quantum matter sector, and such noise is unavoidable in order to resolve paradoxes that arise in semiclassical formulations of the Einstein field equations \cite{Hu.Verdaguer:book:2020,Terno:2412.18213}. In that sense, one may argue that the framework contains an intrinsically quantum gravitational ingredient. If the gravitational-wave sector couples to the ancilla system $E=\bar{Q}\bar{Z}$ [cf.\ Eq.~\eqref{eq:Stinespring.rep.embedding}], the reduced (post-$\Tr_E$) dynamics on $QZ$ can exhibit damping even in vacuum, although the existence and scale of such dissipation depend on the specific interaction and are not implied by the dilation alone. While this does not settle whether the framework should be interpreted as a theory of quantum gravity in the traditional sense, it does sharpen the alternative: the CPCQ theory may be viewed as an operationally restricted sector of an underlying fully quantum theory rather than as an irreducibly hybrid theory.

	\section*{Acknowledgments}
	MH is supported by Grant-in-Aid for Scientific Research (Grant Nos.\ 21H05188, 21H05182, and JP19K03838) from the Ministry of Education, Culture, Sports, Science and Technology (MEXT), Japan.\ SM is supported by the Czech Science Foundation through the Junior Star grant 25-17250M and the Quantum Gravity Unit of the Okinawa Institute of Science and Technology (OIST). DRT thanks the Perimeter Institute for Theoretical Physics for hospitality, and Barbara \v{S}oda and Viqar Husain for discussions. We thank Isaac Layton, Jonathan Oppenheim, and Cesare Tronci for critical comments and useful suggestions. Finally, we thank Samuel Blitz (``Dr.\ Blitz'') for featuring this work in his \href{https://www.youtube.com/watch?v=2mq6KgKvl1I}{Peer Review series on YouTube} and pointing out several typographical errors.

	\bibliographystyle{bibstyle}
	\bibliography{CQreferences}
	
	\appendix
	
	\section{Structure and consistency of hybrid dynamics}\label{app:hybrid.structure.consistency}
	
	\subsection{Mathematical framework} \label{app:ss:mathematical.framework}
	Classical states and observables are phase space functions. More precisely, classical probability distributions $\varrho(z,t)$ on the phase space $\gP$ are either integrable functions or tempered distributions such as $\delta\big(z-z_0(t)\big)$, normalized by
	\begin{equation}
		\int \! \varrho(z,t) \; dz = 1 .
	\end{equation}
	Beyond observables such as $q$ and $p$ and their respective functions, we need linear (multiplicative, differential, integral) operators on phase space functions, subject to appropriate restrictions. Their space is denoted by $\eF_\gP$.
	
	Quantum states $\rho$ are standard trace-one, trace-class operators on the Hilbert space $\eH$ of the quantum system Q, and operators on it belong to some operator space $\eO(\eH)$. Vectors of the CQ Hilbert space can be thought of as $\eH$-valued functions of $z\in\gP$. Thus, we can introduce a map 
	\begin{equation}
		\hat{}\,: \; \eO(\eH) \times \eF_\gP \to \eO(\eH)
	\end{equation}
	that generates the CQ operators. 
	
	A CQ state $\hat\rho(z,t)$ satisfies the normalization condition \eqref{eq:CQ.state.normalization} and generates the reduced classical $\varrho(z,t)$ and quantum $\rho(t)$ states via Eq.~\eqref{red-C} and Eq.~\eqref{red-Q}, respectively.     
	For example, {the state of a hybrid system consisting of classical spatial degrees of freedom} and a quantum spin \textonehalf{} is described by a $2 \! \times \! 2$ matrix of phase space functions that are subject to the constraints discussed in Section~\ref{sec:CQ.dynamics}.
	A product form of a CQ state 
	\begin{equation}
		\hat\rho(z,t) = \rho(z,t) \;\! \varrho(z,t)
	\end{equation}
	that describes a normalized quantum state $\rho(z,t)$  conditioned on the classical system being in the state $z$  {does not imply the} absence of correlations between  {observables of the C and Q subsystems}.
	
	Since the Hilbert space operations are the same for Q and CQ objects, we use the hat $\;\!\hat{}\;\!$ only for CQ states $\hat\rho$ to distinguish them from purely quantum states $\rho$.
	
	\subsection{Consistency requirements} \label{app:ss:consistency.requirements}	
	In this subsection, we summarize the most commonly imposed consistency conditions on hybrid dynamics, as discussed in Refs.~\cite{Boucher.Traschen:PhysRevD:1988,Terno:chapter:2025,Terno:FoundPhys:2006,Gay-Balmaz.Tronci:JGeomMech:2022}. These include formal mathematical requirements necessary to ensure well-defined entities, as well as physical ones motivated by embedding the hybrid model into a broader theoretical framework, such as consistency with current experimental constraints.
	
	The most basic purpose of any CQ hybrid scheme is to obtain predictions about the evolution of the C and Q subsystems and the results of measurements performed on them. To this end, any hybrid scheme must identify the classical and the quantum sectors, as well as to produce
	\begin{enumerate}[label=\Roman*.]
		\item A phase space probability density $\varrho$ that satisfies $\varrho \geqslant 0$ and $\int \! \varrho(z,t) \;\! dz = 1$;
		\item A positive semidefinite density operator $\rho$ that satisfies $\langle A \rangle_{\!\rho} = \Tr[\rho A]$ for operators $A$ on $\eH$.
	\end{enumerate}
	{These are derivable from the state $\hat\rho$ of the combined CQ system and represent the information about the states $\varrho$ and $\rho$ in the C and Q subsystems, respectively.} Each individual sector behaves in the usual way, i.e.,
	\begin{enumerate}[label=\Roman*.]
		\setcounter{enumi}{2}  
		\vspace*{2mm}
		\item For a closed CQ system, if C and Q are uncoupled, then $\varrho$ and $\rho$ evolve according to the standard classical and quantum laws, respectively;
		\vspace*{-1mm}
		\item Classical canonical transformations and quantum unitary transformations are realized on C and Q sectors, respectively (equivariance).  
	\end{enumerate}				
	While the evolution of the CQ system may or may not be unitary, and only one of the Schrödinger or Heisenberg pictures may be accessible, it should
	\begin{enumerate}[label=\Roman*.]
		\vspace{-1mm}
		\setcounter{enumi}{4}  
		\item Be compatible with fundamental physical constraints. \\ In particular:
		\vspace*{-1mm}
		\begin{enumerate}[label=(\alph*),leftmargin=1.45em,itemsep=0.1em]
			\small
			\item Respect the relevant conservation laws whenever they hold for the corresponding purely quantum and purely classical systems; otherwise, quantify and bound any violations arising from CP-induced backreaction;
			\item Respect relativistic causality (i.e., no superluminal communication);
			\item Any hybrid interaction, starting from uncorrelated C and Q, must induce on Q a CP map (CPTP overall; CP trace-nonincreasing per outcome) and preserve all constraints of CPTP quantum dynamics, i.e., no extra power in operations or information-processing tasks;
			\item Obey the second law of thermodynamics.
		\end{enumerate}
		\vspace*{-2mm}
	\end{enumerate}
	In V(a), we deliberately refer to the (approximately) conserved quantities and not the Noether charges, as these may be undefined. 
	
	A hybrid dynamics may introduce backreaction of Q on C only if the CQ correlations are generated dynamically, i.e., 
	\begin{enumerate}[label=\Roman*.]
		\setcounter{enumi}{5}  
		\item The quantum purity $\Tr[\rho^2]$ is not a constant of motion.  
	\end{enumerate}
	Although unitary evolution preserves the purity of the combined CQ state, it may act nontrivially across the C and Q subsystems, thereby generating entanglement. Requirement VI thus represents the minimal necessary condition that allows for nontrivial interactions between the C and Q subsystems.
	
	While the minimal goal is to produce reasonable probability distributions, having access to the Heisenberg picture, i.e., explicit evolution equations for all observable operators, is a desirable feature. In this case, it is reasonable to demand that 
	\begin{enumerate}[label=\Roman*.]
		\setcounter{enumi}{6} 
		\item Differences in the predictions of the quantum and hybrid equations of motion for classically observable quantities vanish in the formal limit $\hbar\to 0$ (provided that any diffusion, decoherence, and backreaction coefficients are co-scaled appropriately).
	\end{enumerate}
	Failure to comply with this form of the correspondence principle leads to a breakdown of the classical limit.

\end{document}